\documentclass[aps,twocolumn,nofootinbib,preprintnumbers,citeautoscript]{revtex4}
\usepackage{amsmath,graphicx,, amssymb, amsthm, times,  float,}

\def\be{\begin{equation}}
\def\ee{\end{equation}}

\def\be{\begin{equation}}
\def\ee{\end{equation}}
\def\cD{\mathcal{D}}
\def\cA{\mathcal{A}}
\def\Msun{M_{\odot}}
\newcommand{\gsim}{\raisebox{-.4ex}{$\stackrel{>}{\scriptstyle \sim}$}}

\begin{document}
\title{Implications of a viscosity bound on black hole accretion }
\smallskip\smallskip %%\smallskip
\author{Aninda Sinha}
\affiliation{Centre for High Energy Physics, Indian Institute of
Science, %\\\it 
Bangalore 560012, India}
%\email{asinha@cts.iisc.ernet.in} 
\author{Banibrata Mukhopadhyay}
\affiliation{Department of Physics, Indian Institute of
Science, %\\\it 
Bangalore 560012, India}
% \email{bm@physics.iisc.ernet.in}

%%\vskip2cm \abstract{ 
\begin{abstract}
  {Motivated by the viscosity bound in gauge/gravity duality, we consider the ratio of shear viscosity 
($\eta$) to entropy density ($s$) in black hole accretion flows. We use both an ideal gas equation of state and the QCD equation of state obtained from lattice for the fluid accreting onto a Kerr black hole. The QCD equation of state is considered since the temperature of accreting matter is expected to approach $10^{12}$K in certain hot flows. We find that in both the cases $\eta/s$ is small only for primordial black holes and several orders of magnitude larger than any known fluid for stellar and supermassive black holes. We show that a lower bound on the mass of 
primordial black holes leads to a lower bound on $\eta/s$ and vice versa. Finally we speculate that the Shakura-Sunyaev viscosity parameter should decrease with increasing density and/or temperatures.}
\end{abstract}
  
\maketitle

\newpage
%\section{Introduction}\label{intro}
Accretion flows around black holes and neutron stars are very hot.
In the geometrically thin (and optically thick, i.e., strong interaction between
dense matter and radiation) regime, the
temperature ($T$) of the flow could be $\sim 10^7$K \cite{ss73}. 
The flows in certain temporal classes of micro-quasar GRS~1915+105 \cite{remillard} are of this type. However, 
in the geometrically thick (and optically thin, i.e., weak interaction 
between tenuous matter and radiation) hotter regime, we expect a two-temperature flow
with the ion temperature ($T_i$) as high as
$7\times 10^{11}$K and the electron temperature being much lower. The supermassive black hole system at the centre of
our galaxy Sgr~A$^*$ \cite{rameshnature}, (most of the temporal classes of) the X-ray binary 
Cyg~X-1 \cite{sl75} are of this type. The latter type of flows is radiatively inefficient. 

%While the latter class of flows is gas pressure dominated, the former one
%exhibits radiatively efficient cooling and hence cooler flows.

In the existing literature, accretion flows 
are modelled using an ideal gas equation of state (EoS) along with radiation.
For hot optically thin flows, the underlying radiation mechanisms
could be synchrotron, inverse-Comptonization of soft photons etc.
The argument behind the choice of ideal gas
is that the interaction energy between the constituent particles is small
compared to their thermal energy. While this could be a viable option for cooler optically thick flows, the temperature of hot 
optically thin flows near the horizon of an
astrophysical black hole is of the order of the crossover temperature, $T_c$  (a temperature at low baryon density when the hadrons melt into a Quark Gluon Plasma (QGP)). At low densities (zero chemical potential), lattice QCD calculations \cite{fodor,peter} suggest a crossover 
from the hadronic state to the QGP state at around $1.5-2\times 10^{12}$K.
At such high temperatures, there is expected to be copious pion production in a strongly interacting system.
%It is however well known that at high temperatures and/or high densities,
%quarks and gluons 
%become deconfined to form a plasma.
 This QGP state is thought to 
have existed a few microseconds after the big bang and is being  extensively studied at the Relativistic Heavy Ion Collider (RHIC) and the Large Hadron Collider (LHC). 
Hence, the choice of ideal gas EoS
seems suspicious for the optically thin flows.
Furthermore, $\sim M_\odot$ primordial black holes \cite{carr} formed at $T\sim 10^{13}$K (quark-hadron 
phase transition era) would indeed accrete QGP matter at high density $\sim 10^{18}$gm/cm$^{3}$. 
So the natural question arises: What EoS do we use for matter at such high temperature and/or high baryon density?
At high temperatures and low baryon density, the answer is provided by lattice QCD. At high temperatures and high baryon density, the lattice methods have proved less useful and one would need to appeal to intuition drawn from models.

Hydrodynamic simulations at the RHIC and LHC suggest a small value for the ratio of shear viscosity ($\eta$) to entropy density ($s$) for the QGP state. Such a small value points towards a strongly interacting matter. String theory arguments and the famous gauge/gravity duality in the form of the AdS/CFT correspondence suggest a lower bound \cite{kss} for $\eta/s$:
\be
\frac{\eta}{s}\geq \frac{\mu}{4\pi} \frac{\hbar}{k_B} \sim 0.08 \mu  \frac{\hbar}{k_B}\,.
\ee
While originally $\mu$ was thought to be unity \cite{kss}, we know now that $\mu<1$ \cite{bms} and quite likely 
non-zero \cite{bek} (for recent reviews, see \cite{teaney}). For example, in the gravity model \cite{mps}, 
$\mu\approx 0.414$.  For the purpose of this work we assume that there is a bound and that $\mu \sim 1$. 
One of the main questions that we wish to answer is: How small is $\eta/s$ in black hole accretion? 
We can anticipate the answer to this question before getting into details. First, $\eta$ is a combination 
of a molecular/physical viscosity and a turbulent viscosity. If the molecular viscosity dominates then 
for temperatures around or above $T_c$, we expect $\eta/s$ to be close to $0.1 \hbar/k_B$. More precisely,
we need to use the QCD EoS. If the turbulent viscosity dominates, then the ratio is 
much higher. However, as we will argue, close to the horizon where gravity effects on the flow are strong, 
or in the early universe when density was high, we expect a small ratio.

%However, the minimum 
%possible mass of an astrophysical black hole in an X-ray binary could be as small as $\sim 5\,M_\odot$ 
%(e.g. GRO~J1655−40), not much more massive
%than a black hole in the quark-hadron transition phase in the early universe. 

%If the flow around such astrophysical black holes is
%optically thick, then the corresponding chemical potential appears large hindering the accreting 
%matter to be from QCD/QGP phase. On the other hand, the chemical potential for optically thin flows 
%is very small allowing to flow to be in the QCD/QGP phase.
%Moreover, accretion
%around supermaasive black holes, whether in the optically thick or thin regime, reveals
%small chemical potential rendering the flow of QCD/QGP matter therein.
%Hence, there is no reason to discard that all the above astrophysical flows to follow the QCD/QGP EoS. 

%\section{The QCD equation of state}\label{QCDeos}

%\section{Advective accretion flows around black holes}

There is a huge literature discussing accretion of optically thin hot 
matter around a Kerr black hole in 
full general relativistic as well as pseudo-Newtonian frameworks (e.g. \cite{marek,gp1,gp2,mukhraj10}).
%{\bf REMOVED Following the analysis in Gammie and Popham \cite{gp1,gp2}, we will consider accretion of matter near a Kerr black hole.}
For the present purpose, we will consider the 
one temperature advection-dominated accretion flow (ADAF) model around a Kerr black hole \cite{gp1,gp2}. The units are  
$G=M=c=1$, with $M$ being the mass of the black hole. In what follows we will work in terms of normalized temperatures where the normalization is done through $m_p c^2/k_B$ with $m_p$ being the mass of a proton. 

\begin{figure} [ht]
\begin{tabular}{c c}
\includegraphics[width=0.22\textwidth,angle=0]{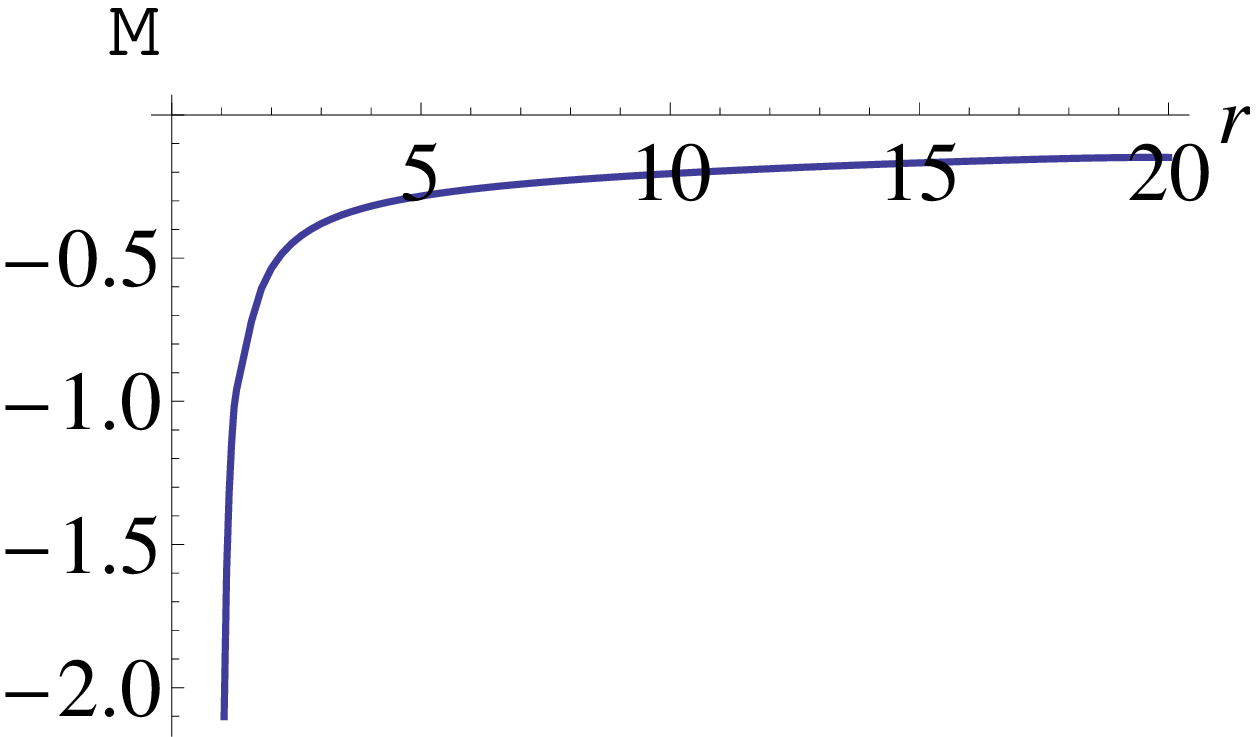} & ~~~~~~ \includegraphics[width=0.22\textwidth,angle=0]{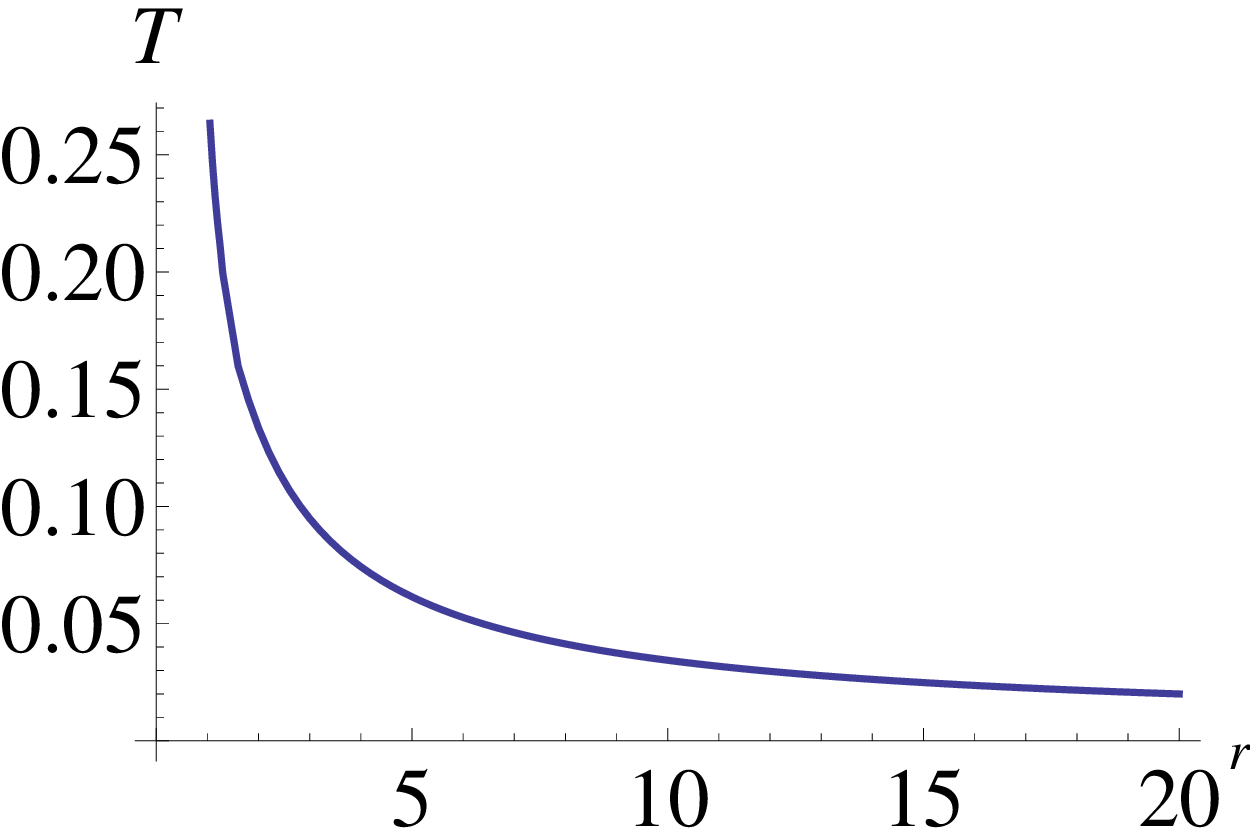} \\
(a) & (b)
\end{tabular}
\begin{tabular}{c}
\includegraphics[width=0.4\textwidth,angle=0]{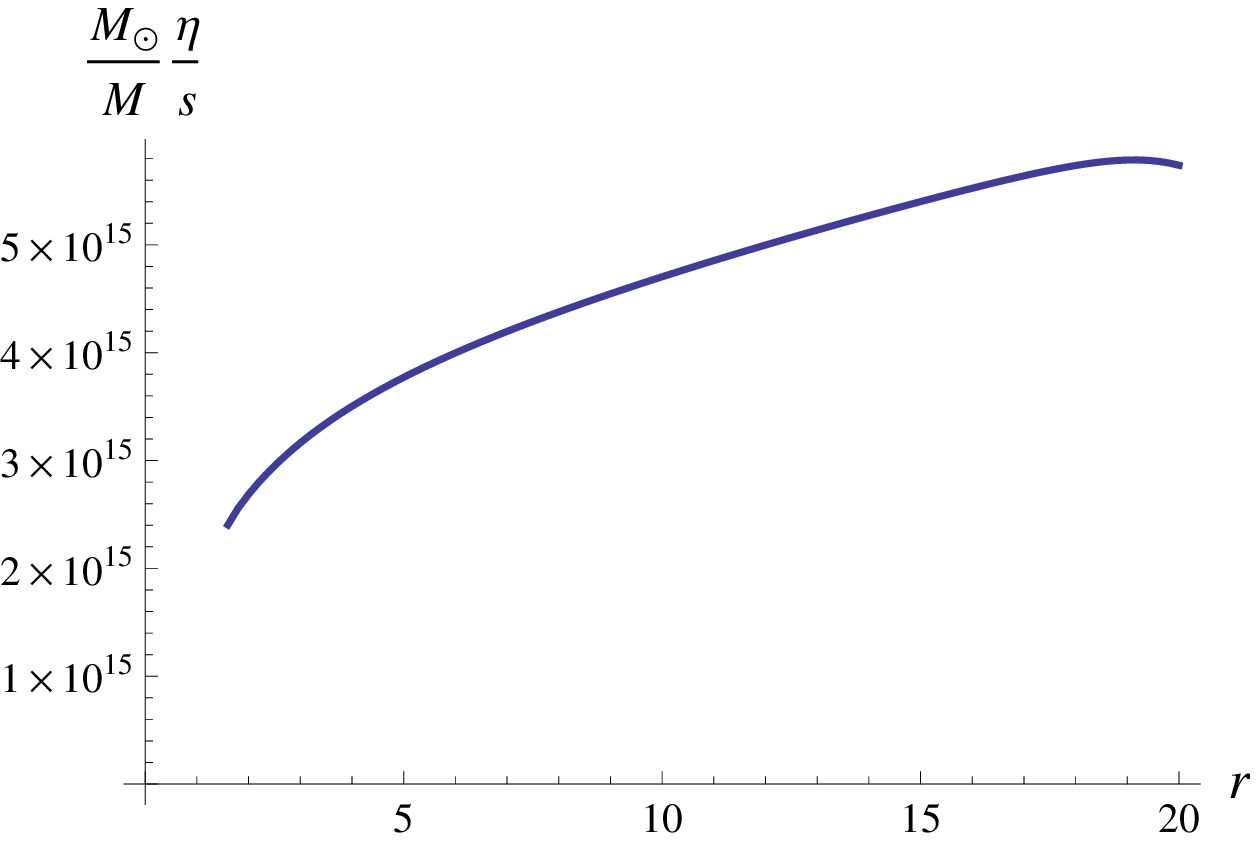}\\
(c)
\end{tabular} 
\caption{Variations of (a) Mach number, (b) temperature, (c) $\eta/s$ in units of $\hbar/k_B$,
as functions of flow radius, when $a=0.999$.  }
\label{figmactrat}
\end{figure}

The equations of continuity and the stress energy conservation satisfy 
\begin{eqnarray}\label{gpeqs}
\nonumber
4\pi r^2 \rho H_\theta \vartheta(\frac{\cD}{1-\vartheta^2})^{1/2}&=&-\dot M\,,\\
\nonumber
 \vartheta(\frac{\cD}{1-\vartheta^2})^{1/2}\left[\partial_r \epsilon -(\epsilon+p)\frac{\partial_r \rho}{\rho}\right]&=&f\Phi\,,\\
\nonumber
\frac{\vartheta}{1-\vartheta^2}\partial_r \vartheta &=&f_r-\frac{1}{\rho\kappa}\partial_r p\,, \\
\dot M \lambda\kappa -4\pi H_\theta r^2 W^r_\phi&=&\dot M j\,.
\end{eqnarray}
Here $r$ is the Boyer-Lindquist radius, $\rho$ the rest-mass density, $H_\theta=H/r$ with $H$ the half
thickness of the flow, $\vartheta$ the radial velocity measured in the co-rotating frame, $\lambda$ the 
specific angular momentum of the flow, $\cD=1-2/r+a^2/r^2$ with 
$a$ being the rotation parameter and $\dot M$ the rest-mass accretion rate. The total energy density 
$\epsilon=\rho+\rho T g(T)$ with $g(T)=(4/(\gamma_0-1)+15 T)/(4+5 T)$ and $\kappa=(\epsilon+p)/\rho$ is the relativistic enthalpy. The function $f_r$ is given by
$$
f_r=-\frac{1}{r^2}\frac{\cA \gamma_\phi^2}{\cD}(1-\frac{\Omega}{\Omega_+})(1-\frac{\Omega}{\Omega_-})\,,
$$
with $\cA=1+a^2/r^2+2 a^2/r^3$, $\gamma_\phi^2=1+\lambda^2(1-\vartheta^2)/(r^2\cA)$ and $\Omega_\pm=\pm(r^{3/2}\pm a)^{-1}$ with $\Omega=2a/(\cA r^3)+\lambda(1-\vartheta^2)^{1/2} \cD^{1/2}/(\gamma_\phi r^2 \cA^{3/2})$ being the angular velocity. When complemented with the gas EoS $p=\rho T$,  the equations (\ref{gpeqs}) enable us 
to solve for $\vartheta, T, \rho, \lambda$ as functions of $r$. In (\ref{gpeqs}), $j$ is a constant and 
should be treated as an eigenvalue for transonic flows. The form of the equations enables us to rescale 
$\rho\rightarrow \dot M \rho$ to set $\dot M=1$ in the numerics. To reinstate physical units, for radial 
velocity we use $\vartheta c$, for specific angular momentum $ \lambda GM/c$, for density $\rho \dot M c^3/G^2 M^2$ 
and for temperature $T m_p c^2/k_B$. 
The viscous stress tensor components entering the equations are $W^r_\phi$ and $\Phi$, given by
\begin{eqnarray}
W^r_\phi&=& F S\,,\nonumber \\
S&=& \frac{\frac{\rho \tau_r (u^r)^2}{F}\frac{d(\lambda\kappa)}{dr}-2\kappa \eta \sigma}{1-\tau_r u^r(2/r+d \ln F/dr)} \,,\nonumber \\
\Phi&=& -2 S \sigma\,,
\end{eqnarray}
with $u^r= \vartheta\cD^{1/2}/(1-\vartheta^2)^{1/2}$.
The expression for $\sigma$ is very lengthy and can be found in the appendix of \cite{gp1}. 
Parametrizing the turbulent viscosity based on the famous Shakura-Sunyaev prescription \cite{ss73} 
and accounting for the relativistic enthalpy factor as in \cite{gp1,gp2} we have
\be
\eta=\alpha \kappa \rho c_s H_\theta r\,.
\ee
Here $\alpha$ is a constant typically taking values $0.01-0.1$. The contribution from the physical (molecular) viscosity is small and the viscosity is turbulence induced. In the present work, we take $\alpha=0.01$ and $\gamma_0=1.4444$.
The quantity 
$\tau_r$ is a relaxation time and is given by $\tau_r=\eta/(\kappa \rho c_s^2)$. At this stage it is a 
phenomenological input. 
The presence of the relativistic enthalpy makes $\eta$ proportional to $\epsilon+p$. 
Using the fact that the entropy density $s=(\epsilon+p)/T\approx \rho (c^2+c_s^2)/T$, we have
\be
\frac{\eta}{s}\approx \kappa \frac{\alpha c_s H_\theta r T}{c^2+c_s^2}\,.
\ee
 Following previous work \cite{gp1,gp2}, 
we reproduce the results using a shooting method shown in FIG. \ref{figmactrat}.
We estimate a rough order of magnitude for $\eta/s$
for $a=0.999$, when the minimum value of $H_\theta$ is of the order of $0.3$ near the horizon 
where $\kappa\sim O(1)$, $c_s\sim 0.4$ and $T\sim 0.3$ for $\alpha=0.01$. This gives us 
\be
\frac{\eta}{s}\approx 4 \times 10^{-4} \frac{GM m_p}{k_B c} \approx 10^{15} \frac{M}{\Msun} \frac{\hbar}{k_B}\,,
\ee
so that for stellar mass black holes, this ratio is of the order of $10^{16}$  in units of 
$\hbar/k_B$,  while for a supermassive black hole this is $>10^{21}$ compared to the QGP value of $0.1$!  
Only for primordial black holes with $M/\Msun \sim 10^{-16}$ will this ratio be comparable to the QGP value. In fact in order to be consistent with the viscosity bound, it appears that 
\be\label{impineq}
M\gsim 10^{-16} \Msun.
\ee
Now, note the following curious fact. Had we been using a larger $\alpha\sim 1$ we would have got $\eta/s\sim 10^{17}M/\Msun \hbar/k_B$. If we use the fact that in the present universe the surviving primordial black holes must have $M>10^{15}$gm we obtain $\eta/s > 0.1 \hbar/k_B$!

%The corresponding plot for $\eta/s$ is shown below.

%\begin{figure} [hb]
%\begin{center}
%\includegraphics[width=0.7\textwidth,angle=0]{ebs} 
%\caption{$\eta/s$ in units of $\hbar/k_B$ against $r$ for $a=0.999$.  }
%\end{center}
%\end{figure}
%\end{center}

FIG. \ref{figmactrat} shows that close to the black hole event horizon $T=0.25$, 
corresponding to $\sim 2.5 \times 10^{12}K$, which is 
above $T_c$. For QGP matter at this temperature we expect a small $\eta/s$.
Hence, the remarkably high value of $\eta/s$ for astrophysical black holes
is quite puzzling. 
Even away from the event horizon, while $T$ decreases (only an order of magnitude) below the crossover temperature, 
it is large enough to question the very large $\eta/s$ we find. 
One possible resolution to this puzzle is that at these 
temperatures the EoS being used is problematic and it is more appropriate to use the 
QCD EoS \cite{peter,fodor}.

The QCD EoS from lattice calculations is specified by the energy density $\epsilon$ and pressure $p$. It is useful to express the result in terms of $I(T)=\epsilon-3p$ which is proportional to the trace anomaly. If we had conformal matter, this quantity would be zero.  At high temperatures ($T\gtrsim 1.97 \times 10^{12}K$), a useful way of fitting the lattice data is given by \cite{peter}
\be
\frac{I(T)}{T^4}=\frac{d_2}{T^2}+\frac{d_4}{T^4}+\frac{c_1}{T^5}+\frac{c_2}{T^{18}}\,,
\ee
where $d_2=0.2405 GeV^2, d_4=0.01355 GeV^4, c_1=-0.0003237 GeV^5, c_2=1.439 \times 10^{-14} GeV^{18}$. A somewhat different fitting function is given in \cite{fodor}. The crossover temperature in \cite{fodor} works out to be around $150\,MeV$ in comparison to what is used in \cite{peter}, which is closer to $190\, MeV$. At low temperatures ($\lesssim 1.97 \times 10^{12}K$), 
\be
\frac{I(T)}{T^{4}}=a_1 T+a_2 T^3+a_3 T^4+a_4 T^{10}\,,
\ee
where $a_1=4.654 GeV^{-1}, a_2=-879 GeV^{-3}, a_3=8081 GeV^{-4}, a_4=-7039000 GeV^{-10}$. The low temperature approximation is found using a matching procedure with the Hadron Resonance Gas model.  From these we find
\be
\frac{p(T)}{T^4}=\int_0^T \frac{dT}{T^5} I(T)\,,
\ee
shown in FIG. 2.
In order to go from $1 MeV$ to the normalized units we have to multiply by $\approx 0.0011$. 
%{\bf NO NEED so that $200 MeV\approx 0.22$ in terms of the normalized temperature.} {\bf CHANGED The variations of energy density ($\epsilon$) and pressure ($p$) are shown in FIG \ref{fight}.}
\begin{figure} [ht]
\begin{tabular}{c c}
\includegraphics[width=0.22\textwidth,angle=0]{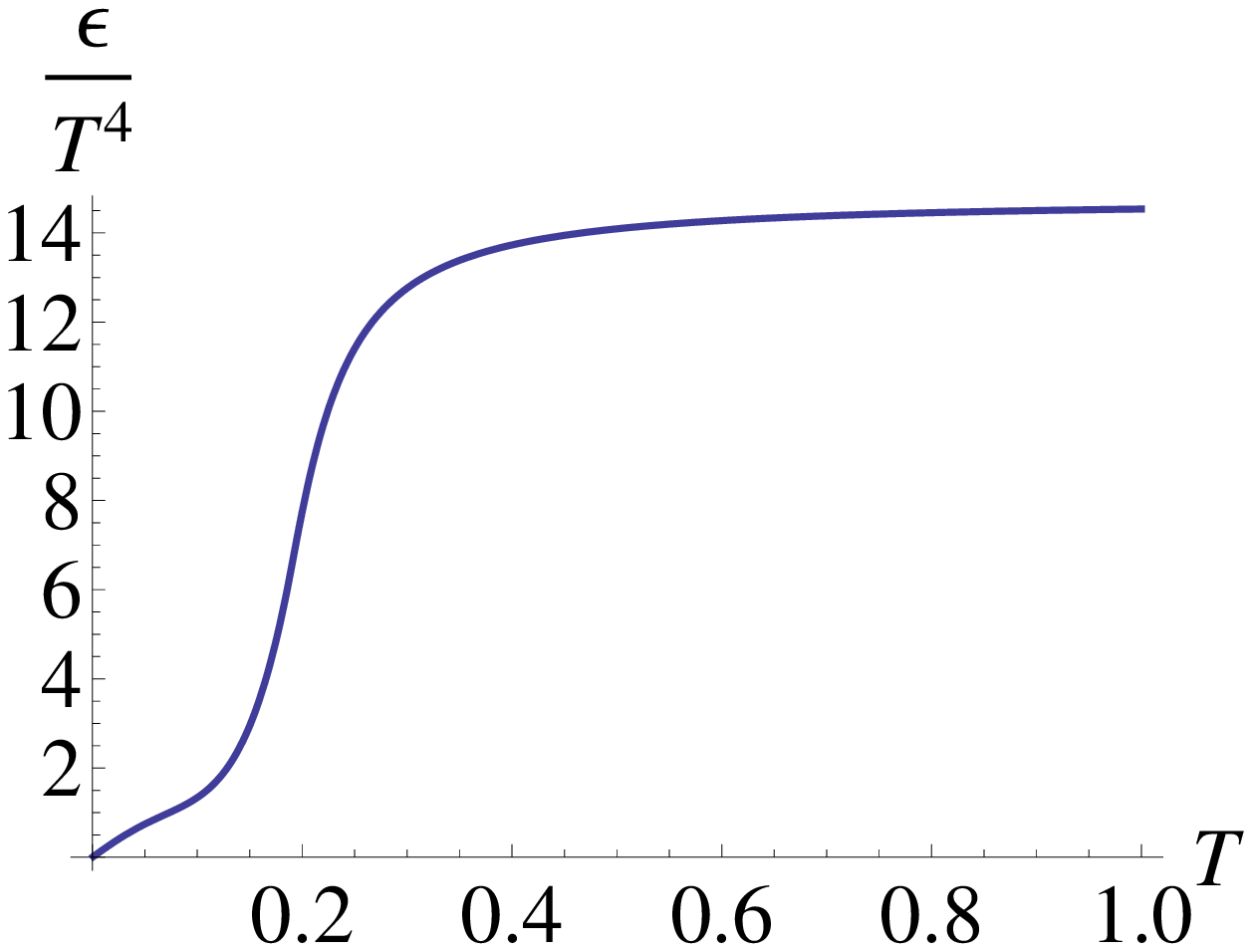} & ~~~~~~ \includegraphics[width=0.22\textwidth,angle=0]{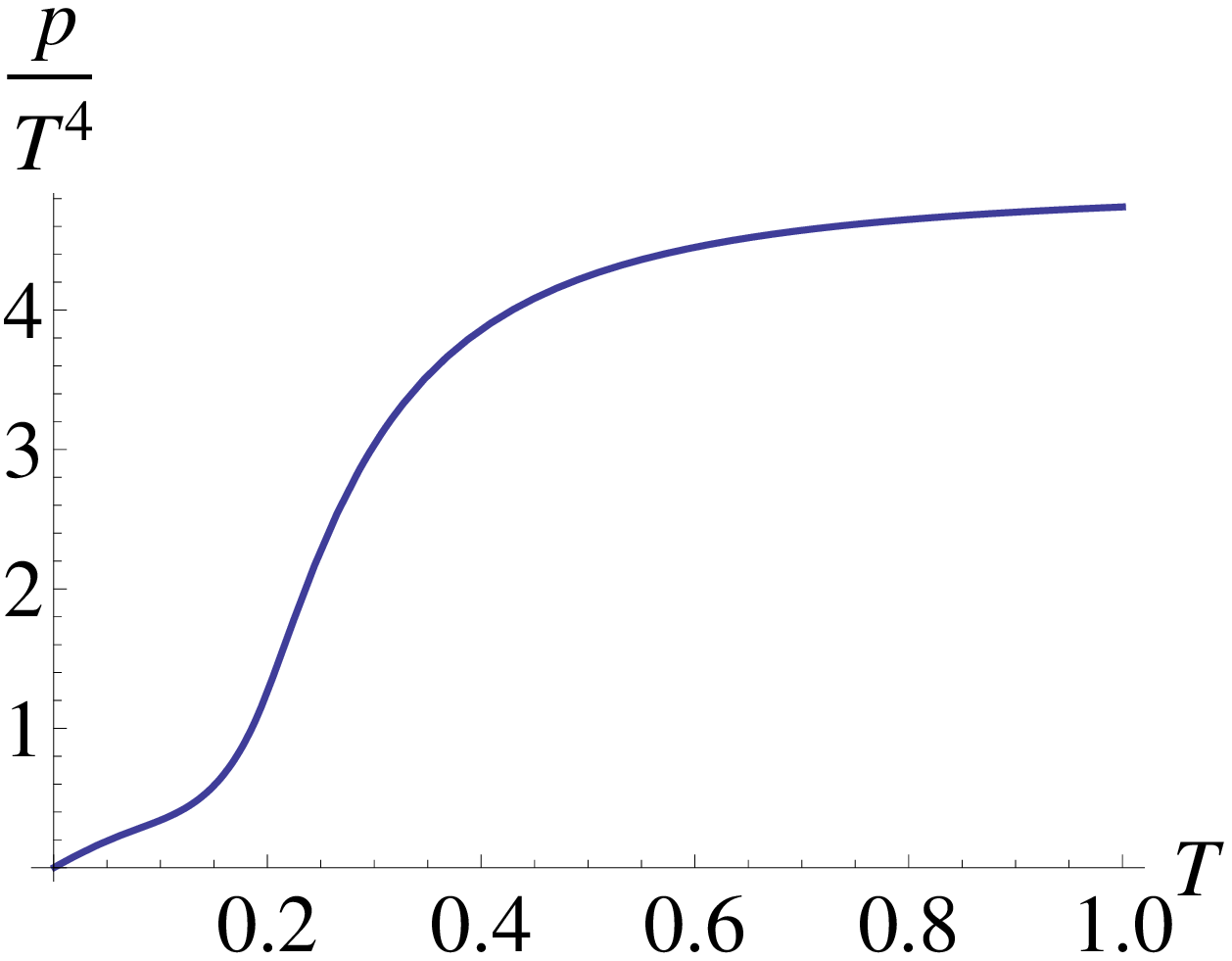} \\
(a) & (b)
\end{tabular}
\caption{Variations of (a) $\epsilon/T^4$, (b) $p/T^4$, as functions of temperature. }
\label{fight}
\end{figure}
%{\bf NO NEED The speed of sound is obtained in the usual way from $c_s^2=\partial p/\partial \epsilon$.} 

 In order to explain the QGP observables for heavy ion flows at RHIC/LHC, it has become apparent that a non-zero shear viscosity, $\eta$, is needed. One way of characterizing $\eta$ which is a dimensionful number, is to take its ratio with the entropy density. It turns out that the RHIC/LHC plasma has \cite{heinz}
\be\label{ebs}
\frac{\eta}{s}\sim 0.1 \frac{\hbar}{k_B}\,.
\ee
Although $\eta$ itself is large (in cgs units $\sim 10^{12}$ gm/cm s), its ratio with $s$ is small. 
To put things into perspective, this ratio for water is around $2$ orders of magnitude greater. 
Since for optically thin accretion flows (e.g. ADAF \cite{ny94}, GAAF \cite{mukhraj10}) the temperature 
near the horizon can be of the order of $10^{12}K$, it is an interesting question to ask if 
$\eta/s$ becomes as small as in equation (\ref{ebs}).

%\section{Advective accretion flows with QCD matter}

%At very high temperatures it is quite likely that the QCD EoS makes better physical sense. 
%It would be interesting to see if there is a QGP crossover from the hadronic phase. 
%This can presumably only take place close to the horizon. 
The QCD EoS (see FIG. 2) cannot be valid 
everywhere since the low temperature behaviours of QCD matter and ordinary matter are quite different,
and far away from the black hole when flow is cooler, the accreting matter will presumably follow the ideal gas EoS. 
Hence, an accurate analysis would involve a knowledge of EoS that is valid everywhere, but this 
appears to be a very hard problem. However, we know that close to the horizon $r_h$ ($r<10 r_h$) flow must be 
transonic. Therefore, we start our computations for the QCD EoS with temperatures at an outer boundary that are similar 
to what we found using the ideal gas 
EoS.
% and ask if the very high temperatures we obtained in the inner region in the case of ideal gas EoS, 
%are still prevalent forthe QCD matter. We also compute the ratio $\eta/s$ throughout the flow. 
%We choose $\alpha=0.01$.

Before we start discussing the solutions, note that  equations (\ref{gpeqs})-(8) exhibit 
the following scaling symmetry:
%\begin{enumerate}
%\item 
(a) Rescaling $\rho$ to $\rho\rightarrow \dot M \rho$ and $j\rightarrow j/\dot M$ leaves the equations invariant. 
%\item 
(b) Rescaling $\epsilon, p$ and $j$ to $\epsilon\rightarrow \beta \epsilon$, $p\rightarrow \beta p$ and $j\rightarrow \beta j$ leaves the equations invariant ($\beta$ being any constant).
%\end{enumerate}
Thus we do not need to carry the actual units of $\epsilon, p, j$. 
%When we convert $\epsilon, p, j$ from the 
%actual physical units to the dimension units (as was done for ideal gas EoS), it turns out that there will be a 
%factor of $0.3 (M/\Msun)^2$. However, in view of the above symmetries, we simply ignore this factor with 
%the understanding that in the values for $\epsilon, p, j$ this factor is present. 
The relaxation time $\tau_r$ is taken to be $\tau_r=b\eta/(s T)$ \cite{roma}. Here $b$ parametrizes our ignorance of the QCD coupling constant, influence of gravity etc. We choose $b=10^{4}$. The choice of $b$ does not alter our conclusions.

\begin{figure} [h]
\begin{tabular}{c c}
\includegraphics[width=0.22\textwidth,angle=0]{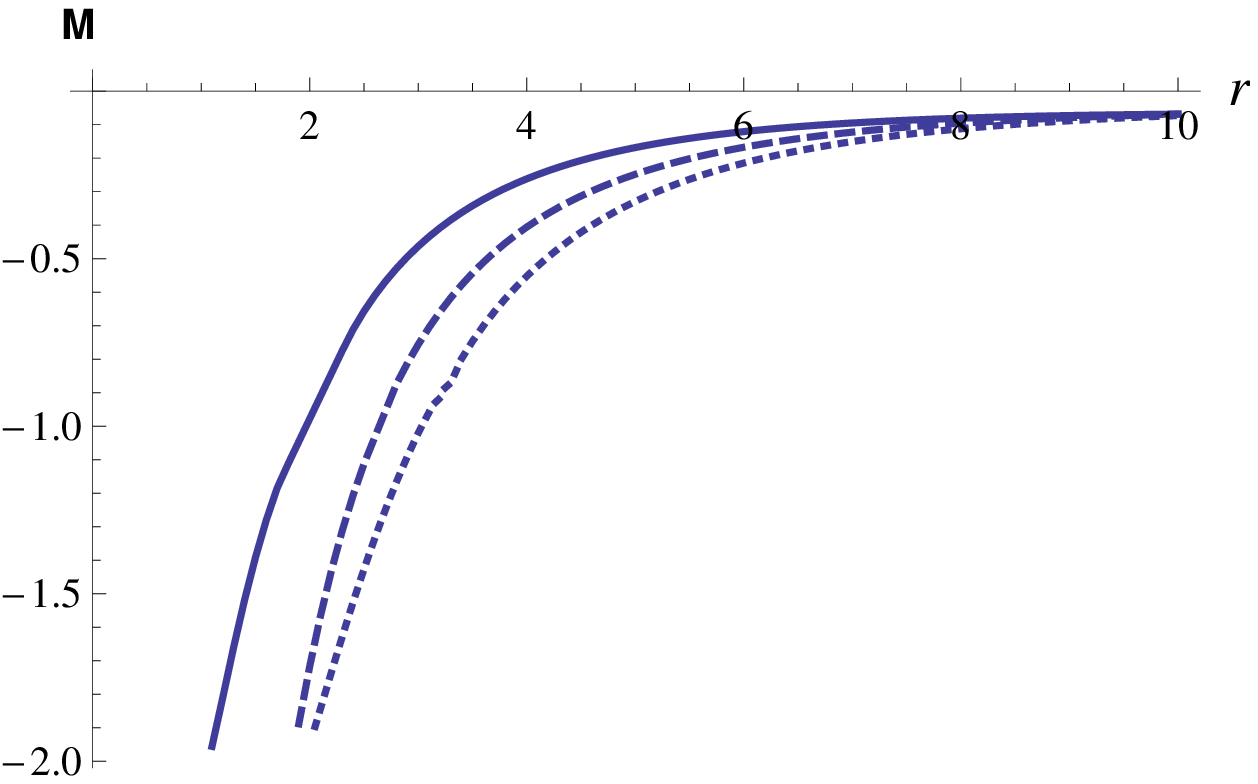} & ~~~~~~ \includegraphics[width=0.22\textwidth,angle=0]{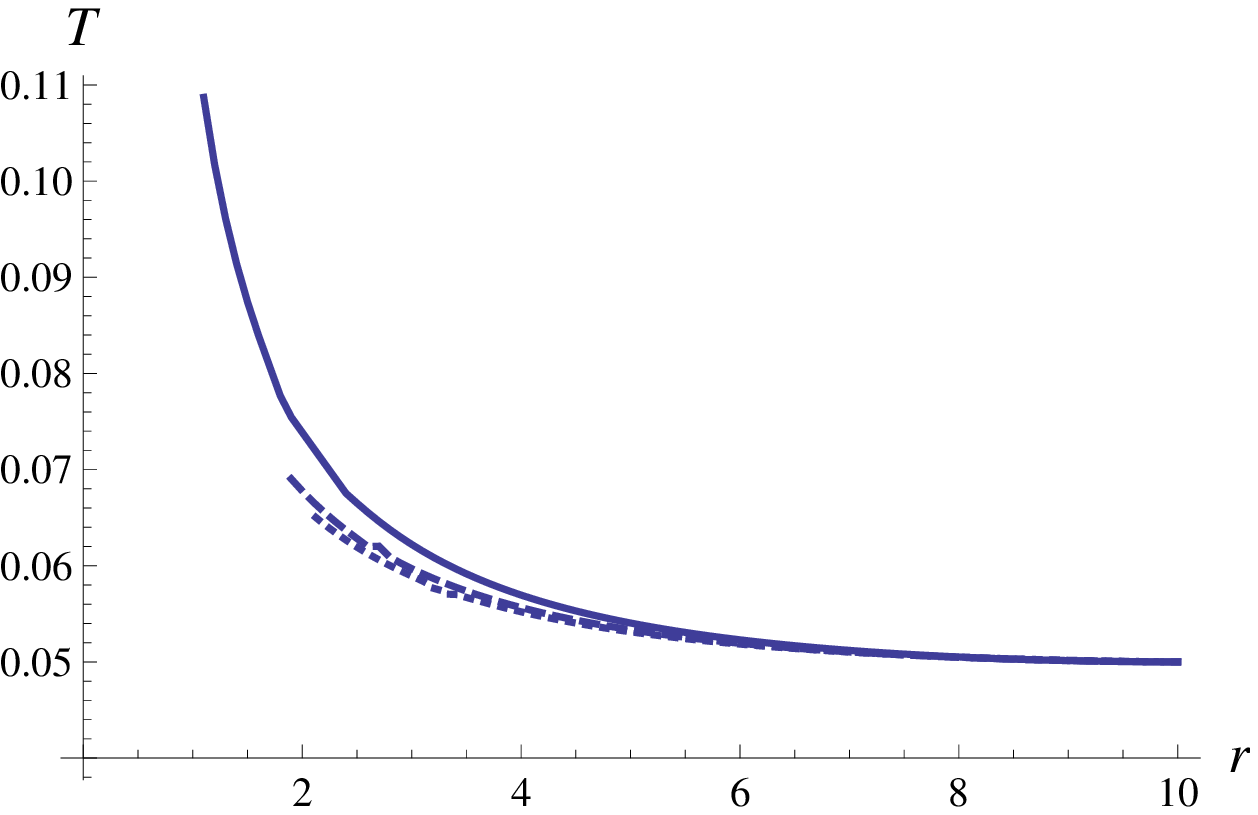} \\
(a) & (b)
\end{tabular}
\caption{Variations of (a) Mach number,  (b) temperature, as functions of flow radius. 
The solid dashed and dotted lines are for $a=0.999, 0.5$ and $0$ respectively.}
\label{qcdsol1}
\end{figure}

 FIG. \ref{qcdsol1} shows the variations of Mach number and temperature as functions of radius. 
%The solid line is for $a=0.999$, the dashed line is for $a=0.5$ while the dotted line is the Schwarzschild case $a=0$.
%{\bf I DON'T IT IS AS IMPORTANT AS TO BE QUOTED In all these cases the eigenvalue $j$ is four orders of magnitude smaller than in the ideal gas case. }
%For $\eta/s$ we find it convenient to define $\delta$ via
%\be
%\dot M =\frac{10^{17}}{\delta} \frac{M}{\Msun} \,\, g s^{-1}\,.
%\ee
%
\begin{figure} [ht]
\begin{center}
\includegraphics[width=0.4\textwidth,angle=0]{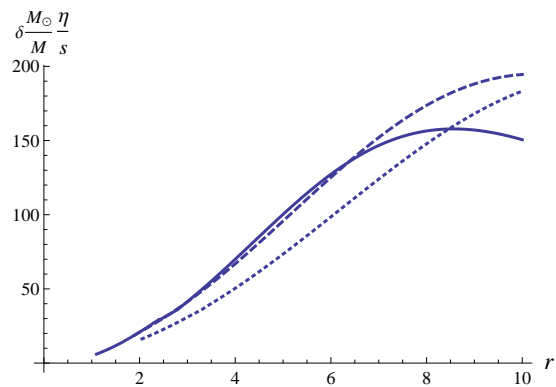} 
\caption{Variations of $\eta/s$ in units of $\hbar/k_B$ as a function of flow radius for the 
cases shown in FIG. \ref{qcdsol1}.  }\label{qcdsol2}
\end{center}
\end{figure}
The variations \cite{hump} of $\eta/s$ for various values of $a$ are shown in FIG. \ref{qcdsol2}. This seems to suggest the same inequality as in equation (\ref{impineq}).

So what have we learnt? In using the QCD EoS we have found the following:
\begin{itemize}
\item The temperature close to $r_h$ never reaches as high as what the ideal gas EoS dictates.
\item The value of $\eta/s$ is small only for primordial black holes 
and gives a result that is quite similar 
to what was found using the ideal gas EoS, for all black holes.
\item Using the fact that the mass of primordial black holes in the present universe must be greater than $10^{15}$gm, we find that $\eta/s>0.1 \alpha\hbar/k_B$. Although for $\alpha<1$, $\eta/s<\frac{1}{4\pi}\hbar/k_B$, this is perfectly consistent with the fact that the original KSS bound can in fact be violated in holography\cite{bms}.
\end{itemize}

Let us conclude with some speculations. 
 In order to apply the viscosity bound for black holes of mass $0.1 \Msun$ in the early 
universe around $T_c$ (which is in the radiation dominated era), one must consider an optically thick 
flow model \cite{ss73,nt73}, unlike the one we considered previously. In this quark-hadron
phase transition era, the black holes must 
be accreting QGP matter. However, here we face the following 
conundrum. First, for QGP matter, the Reynolds number ($Re$) is $O(1)$. Hence, we do not expect 
turbulent viscosity to dominate in this era and we expect a small $\eta/s$. However, if we take 
$\alpha\sim 0.01$ then $\eta/s$ comes out to be $\sim O(10^9)\hbar/k_B$ in the inner accreting region
in our computations. Second, the temperature in the inner region in this model 
\cite{ss73,nt73} for $\alpha\sim 0.01-0.1$ is only around $10^7-10^8$K. To get  $T>10^{11}$K for 
$\dot M=\xi \dot M_{Edd}$ \cite{medd}, the value of $\alpha$ should be very small ($ < 10^{-14}$) 
which in turn leads to a small $\eta/s \sim \xi^2 \hbar/k_B$ with $Re\sim 10$. Such a low $\alpha$ 
also leads to a large density which is consistent with the fact that the universe was 
very dense in this era. This seems to suggest that $\alpha$ should decrease with increasing density. 
%One may argue that the large chemical potential in this era may make the ratio larger than this estimate. But AdS/CFT intuitions \cite{sinha} suggest that increasing the baryon density
%will only lower the value of $\eta/s$.
Now, does it make sense to have $\eta/s> O(10^{15})\hbar/k_B$ near $T_c$, which appears to be the case 
for stellar and supermassive black holes discussed above? Near the black hole, we expect gravity to make the 
flow sub-Keplerian and quasi-spherical, which is more ``ordered", diminishing turbulence. This is supported 
by the fact that $Re$ for QGP and the optically thin disk flows are comparable and as such if matter was 
in the form of QGP,  turbulent viscosity would not dominate. A more natural solution is to demand that 
$\alpha$ becomes small near $T_c$. Thus our conjecture is the following:

{\it The Shakura-Sunyaev viscosity parameter $\alpha$ should be a function of temperature and density such that it decreases with increasing temperature and/or density. However,  low $\alpha$ would only support  
sub-Keplerian optically thin flows. Therefore, the inner accretion is necessarily sub-Keplerian.}

% It will be interesting to explore the ramifications of this further.

 Finally let us point out a somewhat worrying aspect to having a large $\eta/s$. First, we note that for the hydrodynamic 
approach to be valid, it had better be true that the stress tensor is treated as an expansion in 
gradients of the velocity. In particular 
$
T^{\mu\nu}=p g^{\mu \nu}+\rho \kappa u^\mu u^\nu+ t^{\mu\nu}\,,
$
where $t^{\mu\nu}$ is proportional to $\eta$ and involves gradients of the velocity. We have to ensure that the first order terms, $t^{\mu\nu}$, are much smaller than the zeroth order 
terms, $p g^{\mu \nu}, \rho \kappa u^\mu u^\nu$. When we compute the ratio of these two terms along 
the flow, it turns out that the contribution from $t^{\mu\nu}$ is around $20-25\%$ of that of the leading term. 
Hence, the numerics appear to be suspect since the subleading terms in the gradient expansion will be important. It may be worth investigating effects such as cavitation \cite{kls}  in this context. It seems that lowering the viscosity parameter along the lines we have hinted at above may ameliorate this problem.

{\bf Acknowledgments} : We thank Z. Fodor for useful correspondence and R. Narayan and  C. Gammie for comments on the draft. 
Thanks are due to H. K. Aravind, S. Bhat Jr., A. Bhattacharyya, 
R. Misra for many discussions. This work was partly supported by an ISRO grant ISRO/RES/2/367/10-11 (BM)
and a Ramanujan fellowship (AS), funded by government of India.


\begin{thebibliography}{99}
\bibitem{ss73}
N. I. Shakura, R. A. Sunyaev, %``Black holes in binary systems. Observational appearance,''
A\&A {\bf 24}, 337 (1973).

\bibitem{remillard}
R. A. Remillard, J. E. McClintock, ARA\&A {\bf 44}, 49 (2006).

\bibitem{rameshnature}
R. Narayan, I. Yi, R. Mahadevan, Nature {\bf 374} 623 (1994).

\bibitem{sl75}
A. P. Lightman, S. L. Shapiro, ApJ {\bf 198}, 73 (1975).

%\bibitem{medd} $\dot M_{Edd}$ is the Eddington accretion rate defined to be $10^{18} M/\Msun$ gm/s.
\bibitem{fodor}
%  S.~Borsanyi, G.~Endrodi, Z.~Fodor, A.~Jakovac, S.~D.~Katz, S.~Krieg, C.~Ratti, K.~K.~Szabo,
  S.~Borsanyi et al.
  %``The QCD equation of state with dynamical quarks,''
  JHEP {\bf 1011}, 077 (2010).
 % [arXiv:1007.2580 [hep-lat]].

\bibitem{peter}
  P.~Huovinen, P.~Petreczky,
  %``QCD Equation of State and Hadron Resonance Gas,''
  Nucl.\ Phys.\  {\bf A837}, 26-53 (2010).
%  [arXiv:0912.2541 [hep-ph]].

\bibitem{carr}
  B.~J.~Carr,
  %``Primordial black holes: Do they exist and are they useful?,''
  arXiv:astro-ph/0511743.

\bibitem{kss}
  P.~Kovtun, D.~T.~Son, A.~O.~Starinets,
  %``Viscosity in strongly interacting quantum field theories from black hole physics,''
  Phys.\ Rev.\ Lett.\  {\bf 94}, 111601 (2005).
 % [hep-th/0405231].

\bibitem{bms}Y.~Kats and P.~Petrov,
  %``Effect of curvature squared corrections in AdS on the viscosity of the dual gauge theory,''  
JHEP {\bf 0901}, 044 (2009) \\% [arXiv:0712.0743 [hep-th]].
 M.~Brigante, H.~Liu, R.~C.~Myers, S.~Shenker and S.~Yaida,
  %``The Viscosity Bound and Causality Violation,'' 
 Phys.\ Rev.\ Lett.\  {\bf 100}, 191601 (2008)\\  %[arXiv:0802.3318 [hep-th]].
  A.~Buchel, R.~C.~Myers, A.~Sinha,
  %``Beyond eta/s = 1/4 pi,''
  JHEP {\bf 0903}, 084 (2009).
 % [arXiv:0812.2521 [hep-th]].

\bibitem{bek}
  I.~Fouxon, G.~Betschart, J.~D.~Bekenstein,
  %``The Bound on viscosity and the generalized second law of thermodynamics,''
  Phys.\ Rev.\  {\bf D77}, 024016 (2008).

\bibitem{teaney}
  T.~Schafer, D.~Teaney,
  %``Nearly Perfect Fluidity: From Cold Atomic Gases to Hot Quark Gluon Plasmas,''
  Rept.\ Prog.\ Phys.\  {\bf 72}, 126001 (2009).
 % [arXiv:0904.3107 [hep-ph]]. \\
  A.~Sinha, R.~C.~Myers,
  %``The Viscosity bound in string theory,''
  Nucl.\ Phys.\  {\bf A830}, 295C-298C (2009).
 % [arXiv:0907.4798 [hep-th]].\\
  S.~Cremonini,
  %``The Shear Viscosity to Entropy Ratio: A Status Report,''
  arXiv:1108.0677 .


\bibitem{mps}
  R.~C.~Myers, M.~F.~Paulos, A.~Sinha,
  %``Holographic studies of quasi-topological gravity,''
  JHEP {\bf 1008}, 035 (2010).
 % [arXiv:1004.2055 [hep-th]].

\bibitem{marek}
  M.~A.~Abramowicz,  P.~C.~Fragile,
  %``Black Hole Accretion Disks,''
  arXiv:1104.5499.
%\cite{Gammie:1997ct}
\bibitem{gp1}
  C.~F.~Gammie, R.~Popham,
  %``Advection dominated accretion flows in the Kerr metric: 1. Basic equations,''
  %Submitted to: Astrophys.J..
 ApJ {\bf 498}, 313 (1998). 


\bibitem{gp2}
  R.~Popham, C.~F.~Gammie,
 % ``Advection dominated accretion flows in the kerr metric. 2. Steady state global solutions,''
ApJ {\bf 504}, 419 (1998).

\bibitem{heinz} 
  U.~W.~Heinz, C.~Shen and H.~Song,
  %``The viscosity of quark-gluon plasma at RHIC and the LHC,'' 
 arXiv:1108.5323 [nucl-th].

\bibitem{mukhraj10} S. R. Rajesh, B. Mukhopadhyay, MNRAS {\bf 402}, 961 (2010).

%\bibitem{commal}
%Had we taken $\alpha=0.1$ then $\eta/s \sim 10^{16}M/\Msun \hbar/k_B$.

\bibitem{ny94}
R. Narayan, I. Yi, ApJ {\bf 428}, L13 (1994).

\bibitem{roma}
  P.~Romatschke,
  %``New Developments in Relativistic Viscous Hydrodynamics,''
  Int.\ J.\ Mod.\ Phys.\  {\bf E19}, 1-53 (2010).
  %[arXiv:0902.3663 [hep-ph]].

\bibitem{hump} The hump in the case of $a=0.999$ is an artifact of our choice of initial conditions and will be absent in a more complete analysis. 



%\bibitem{bmas}
%B.~Mukhopadhyay and A.~Sinha, in preparation.

\bibitem{nt73}
I. D. Novikov, K. S. Thorne, %``Astrophysics of black holes,''
{\it Black holes (Les astres occlus)}, 343 (1973).







\bibitem{medd} $\dot M_{Edd}$ is the Eddington accretion rate defined to be $10^{18} M/\Msun$ gm/s.

%\bibitem{sinha}
 % R.~C.~Myers, M.~F.~Paulos, A.~Sinha,
  %``Holographic Hydrodynamics with a Chemical Potential,''
 % JHEP {\bf 0906}, 006 (2009).
  %[arXiv:0903.2834 [hep-th]]

\bibitem{kls}
  A.~Klimek, L.~Leblond, A.~Sinha,
  %``Cavitation in holographic sQGP,''
  Phys.\ Lett.\  {\bf B701}, 144-150 (2011).
  %[arXiv:1103.3987 [hep-th]].

\end{thebibliography}
\end{document}